\documentclass[12pt]{article}

\usepackage{amssymb,hyperref,latexsym}

\makeatletter \@addtoreset{equation}{section} \makeatother
\renewcommand{\theequation}{\thesection.\arabic{equation}}
\newcounter{parentequation}
\newenvironment{subequations}{%
  \refstepcounter{equation}%
  \begingroup
\let\protect\noexpand
  \edef\@tempa{\def\noexpand\theparentequation{\theequation}}%
  \expandafter
  \endgroup\@tempa
  \setcounter{parentequation}{\value{equation}}%
  \setcounter{equation}{0}%
  \def\theequation{\theparentequation\alph{equation}}%
  \ignorespaces
}{%
  \setcounter{equation}{\value{parentequation}}%
}
\def\al{\alpha}
\def\be{\beta}
\def\ga{\gamma}
\def\albar{\bar{\alpha}}
\def\bebar{\bar{\beta}}
\def\gabar{\bar{\gamma}}

\begin{document}

\thispagestyle{empty}

\begin{flushright}
\hfill{HU-EP-02/49} \\
\hfill{hep-th/0211118}
\end{flushright}

\vspace{15pt}

\begin{center}{ \LARGE{\bf
Non-K\"ahler String Backgrounds \\[4mm]
and their Five Torsion Classes}}

\vspace{40pt}

{\bf G. L. Cardoso, G. Curio, G. Dall'Agata, D. L\"ust}

\vspace{10pt}

{\it  Humboldt Universit\"at zu Berlin,
Institut f\"ur Physik,\\
Invalidenstrasse 110, 10115 Berlin,
Germany}\\[1mm]
{E-mail: gcardoso,curio,dallagat,luest@physik.hu-berlin.de}

\vspace{10pt}

and

\vspace{10pt}

{\bf 
P. Manousselis and G. Zoupanos}

\vspace{10pt}

{\it Physics Department, National Technical University,\\
Zografou Campus, 157 80 Athens, Greece}
\\[1mm]
{E-mail: pman@central.ntua.gr, George.Zoupanos@cern.ch}

\vspace{50pt}

{ABSTRACT}

\end{center}

We discuss the mathematical properties of six--dimensional
non--K\"ahler manifolds which occur in the context of ${\cal N}=1$
supersymmetric heterotic and type IIA string compactifications with
non--vanishing background H--field.  The intrinsic torsion of the
associated $SU(3)$ structures falls into five different classes.
For heterotic compactifications we present an explicit dictionary
between the supersymmetry conditions and these five torsion classes.
We show that the non--Ricci flat Iwasawa manifold solves the
supersymmetry conditions with non--zero H--field, so that it is a
consistent heterotic supersymmetric groundstate.

\newpage

\section{Introduction}

Superstring compactifications on internal spaces with non--vanishing
expectation values for additional `matter' fields are interesting for
various reasons.
In particular, they offer new prospects for obtaining more realistic
string models capable of describing (part of the) four--dimensional
spacetime physics.
For instance, in superstring compactifications with non--vanishing
H--field fluxes (see eg.  \cite{Strominger:1986uh}--
\cite{D'Auria:2002th}) a potential gets generated that results in the
lifting of some of the flat directions, i.e. to a fixing of the value
of some of the geometric moduli of the underlying compact internal
space.
Moreover, in the context of intersecting brane world models with
D6--branes wrapped around internal 3--cycles, it is possible to
construct standard model like spectra in a systematic way (see eg. 
\cite{Ibanez:2001dj,Blumenhagen:2002vp} and other references therein).
Warped compactification with fluxes and/or branes may also explain the
origin of various hierarchies in string theory
\cite{Randall:1999vf}--
\cite{Frey:2002hf},\cite{Giddings:2001yu}.

Spacetime supersymmetry imposes strong restrictions on the allowed
form of the string background fields.
For instance, it is well known that in order to obtain purely
geometric compactifications (without any further background
expectation values) with either ${\cal N}=1$ spacetime supersymmetry
in the heterotic context or ${\cal N}=2$ spacetime supersymmetry in
the type II context, the internal space must be a complex K\"ahler
manifold with vanishing Ricci tensor, i.e. a Calabi--Yau space
\cite{Candelas:1985en}.
However, when turning on additional background fields, which may
correspond to the presence of branes, of fluxes, of a non--constant
dilaton and/or of a non--trivial warp factor, the internal geometry
will not remain Ricci--flat anylonger, and may not even remain
complex.
Since in this more generic situation the tools of complex geometry do
not apply, the mathematical structure of these spaces is much less
understood than in the case of Calabi--Yau spaces.

Heterotic string compactifications on an internal space with
non--trivial warp factor, dilaton and H--field background were first
discussed in \cite{Strominger:1986uh}.
Related compactifications in type I, type II and eleven--dimensional
supergravity were also investigated in \cite{deWit:1987xg}.
The requirement of ${\cal N}=1$ spacetime supersymmetry forces the
internal six--dimensional space to come equipped with a certain
$SU(3)$ connection with torsion, 
provided by the H--field \cite{Strominger:1986uh}.
It turns out that certain properties of six--dimensional manifolds
with torsion and $SU(3)$ structure have been systematically worked out
in the mathematical literature \cite{fried}.
It has been shown \cite{chiossi} that the allowed torsion tensors have
to fall into five different 
classes\footnote{The 
mathematical torsion is $\nabla J$, and it is related to the
$H$--field by $\nabla_m J_n\,^p = 
H_{sm}{}^p {J_n}^s + {H^s}_{mn} 
{J_s}^p $.}.
The resulting manifolds are quite restricted and may or may not be
complex.
One aim of the present paper is to provide an explicit dictionary
between the mathematical classification of torsion of compact
six--dimensional spaces with $SU(3)$ structure and the conditions
leading to ${\cal N}=1$ supersymmetric backgrounds in heterotic string
theory.

The same mathematical classification of torsion may also be used in
the context of type IIA compactifications on six--dimensional spaces
$M$ with additional H--field backgrounds or wrapped D--branes.
Again the gravitational backreaction of the fluxes or the branes is so
strong that the internal type IIA geometry ceases to be of the
Calabi--Yau type.
One nice way to understand this feature is provided by the lift of
type IIA theory to eleven--dimensional M--theory.
Let us consider the case when, upon lifting to eleven dimensions, the
type IIA H--fluxes become completely geometrized, that is to say that
they completely originate from the eleven--dimensional metric.
This is, for instance, the case for the background of the Ramond
one--form gauge potential, as can be seen by a generalized Hopf
fibration procedure.
Similarly, it is well known (see e.g. \cite{Townsend:1997wg}) that the
dilaton and H--field background of wrapped D6--branes become purely
geometrical in eleven dimensions, namely the D6--branes correspond to
the metric of the eleven--dimensional Eguchi--Hanson instanton.
In all these purely geometric M--theory compactifications with ${\cal
N}=1$ supersymmetry in four spacetime dimensions it is known that the
internal space must be a seven--dimensional manifold with $G_2$
holonomy.  By constructing the $G_2$ space as a fibration over the
six--dimensional base $M$, one can deduce the corresponding $SU(3)$
structure of $M$ in type IIA from the associated $G_2$ structure in
M--theory.
Conversely, one can give conditions on the torsion of $M$ which allow
for an uplift of $M$ to a seven--dimensional $G_2$ manifold
\cite{chiossi}.

The main results of this paper can be summarised as follows.
We give an extensive dictionary between the known mathematical
classification of torsion associated with compact six--dimensional
manifolds with $SU(3)$ structure and the necessary and sufficient
conditions derived in \cite{Strominger:1986uh} for four--dimensional
${\cal N}=1$ spacetime supersymmetry in heterotic string
compactifications.
Since spacetime supersymmetry requires the six--dimensional manifolds
to be complex, the associated torsion $\tau$ has to lie in
\begin{equation}
\label{w3w4w5}
\tau \in {\cal W}_3 \oplus {\cal W}_4 \oplus {\cal W}_5\,.
\end{equation}
In addition, spacetime supersymmetry requires that
\begin{equation}
2\, {\cal W}_4 + {\cal W}_5 = 0 
\label{w4w5}
\end{equation} 
with both 
\begin{equation}
{\cal W}_4\;\;\; {\rm and} \;\;\; {\cal W}_5 \;\;\; \hbox {exact and real} \;. 
\label{w4w5ex}
\end{equation}
The presence of torsion results in the 
introduction of a generalized spin connection.
We will denote the associated 
generalized Riemann tensor by ${\tilde R}_{mnpq}$.
Spacetime supersymmetry then also implies the vanishing of the generalized
Ricci two--form
\begin{equation}
{\tilde R}_{mn pq} \, J^{pq} =0 \;.
\end{equation}
The associated generalized Ricci tensor ${\tilde R}_{nq}$, however, need
{\it not} vanish.

An example of a six--dimensional manifold satisfying (\ref{w3w4w5}),
(\ref{w4w5}) and (\ref{w4w5ex}) is given by the Iwasawa manifold. 
This manifold was, to a certain extent, already considered in
\cite{Strominger:1986uh}.
The moduli space of its complex structures consists of two
disconnected components.  
For any choice of a complex structure one
finds that ${\cal W}_4 = {\cal W}_5 =0$.
Explicit computation of the associated generalized Ricci two--form and
of the generalized Ricci tensor shows that while the former vanishes,
the latter doesn't.
In order for the Iwasawa manifold to be a consistent background for
the heterotic string, certain additional conditions need to be met in
the gauge sector.
We show that the Bianchi identity for the three--form $H$ is
non--trivial and that it may be solved by turning on an abelian
background gauge field.
We note that this way of solving the Bianchi identity is very
different from the usual procedure of embedding the spin in the gauge
and of the associated construction of $SU(3)$ instantons.

The Iwasawa manifold possesses ${\cal W}_4 = {\cal W}_5=0$.  One may
wonder whether it is possible to construct complex manifolds with 
${\cal W}_4 \neq 0, {\cal W}_5 \neq 0$ satisfying (\ref{w4w5}) and 
(\ref{w4w5ex}).
A possible technique for finding such spaces is given by
the following.  
Suppose that one has constructed a complex manifold $P$ for 
which both ${\cal W}_4$ and ${\cal W}_5$ are exact and real, so that
${\cal W}_4 = d \Lambda,  {\cal W}_5 = d \Sigma$, and hence
$2 {\cal W}_4 + {\cal W}_5 = d (2 \Lambda + \Sigma) \equiv d f$.
Then, by performing the following rescaling of the metric, $g \rightarrow 
{\rm e}^{2f} g$, it can be shown that $2 {\cal W}_4 + {\cal W}_5 - df =0 
\rightarrow 2 {\cal W}_4 + {\cal W}_5 =0$.  
Therefore, the complex manifold $P$ has a chance of providing a
consistent background for the heterotic string.

The paper is organized as follows.  
In section 2 we give a review of the mathematical classification of
the intrinsic torsion of $SU(3)$ structures in terms of five classes
as well as of the derivation of $G_2$ structures from $SU(3)$
structures \cite{chiossi}.
In section 3 we present the dictionary between the mathematical
classification of torsion and the necessary and sufficient conditions
derived in \cite{Strominger:1986uh} for four--dimensional ${\cal N}=1$
spacetime supersymmetry in heterotic string compactifications.
In section 4 we show that the Iwasawa manifold provides a consistent
background for the propagation of the heterotic string.
Section 5 contains an outlook.
Appendix A contains a review on dynamical and fibered $G_2$ structures and
their relation to type IIA compactifications.
Appendix B contains a review of the derivation of the conditions
for ${\cal N}=1$ spacetime supersymmetry obtained in \cite{Strominger:1986uh}.

\section{$SU(3)$ structures and the five classes of torsion}
\label{mathpart}

In this section we review \cite{chiossi} the classification of the
intrinsic torsion of $SU(3)$ structures in terms of five classes, as
well as the derivation of $G_2$ structures from $SU(3)$ structures.

We consider a six--dimensional manifold $M$ equipped with a Riemannian
metric $g$ and an almost--complex structure $J$, with its associated
two--form\footnote{In the following, by an abuse of notation, we will
use the same symbol $J$ for both the almost--complex structure and its
associated two--form.}.
At least locally one can choose a local orthonormal basis $(e^1 ,\ldots , e^6)$ such 
that the almost complex structure is 
\begin{equation}
J = e^1 \wedge e^2 + e^3 \wedge e^4 + e^5 \wedge e^6\,.
\label{Jdef}
\end{equation}
It follows that 
a basis of $(1,0)$--forms is given by
\begin{equation}
e^i +  i\,ÊJ \cdot e^{i} \in \Lambda^{(1,0)}\,, 
\label{10form}
\end{equation}
where $J \cdot e^a = J^a\,_b \,e^b $
and consequently $J 
\cdot J = - 1$.
The above defines a $U(3)$ structure. 
An $SU(3)$ structure, on the other hand, is determined by the $(3,0)$--form 
\begin{equation}
\Psi = \left(e^1 + {\rm i} e^2\right) \wedge \left(e^3 + {\rm i} e^4\right) 
\wedge \left(e^5 + {\rm i} e^6\right) \,.
\label{Psi}
\end{equation}
This form has norm 1 and it is subject to the compatibility 
relations $J \wedge \psi_{\pm}  =  0$, $\psi_{+} \wedge \psi_{-}  =  
\frac23 J \wedge J \wedge J $,
where we used the further split into real and imaginary parts
\begin{equation}
\Psi = \psi_{+} + {\rm i} \, \psi_-\,,
\label{psipm}
\end{equation}
with $\psi_- = J \cdot \psi_{+}$.

The failure of the holonomy group of the Riemann connection of $g$ to 
reduce to $SU(3)$ can be measured by the intrinsic torsion $\tau$, 
which is identified with $\nabla J$, if $\nabla$ is the covariant 
derivative with respect to the Riemannian connection.
The space to which the torsion belongs can be decomposed into five 
classes \cite{chiossi}
\begin{equation}
\tau \in {\cal W}_1 \oplus {\cal W}_2 \oplus {\cal W}_3 \oplus {\cal 
W}_4 \oplus {\cal W}_5  \,,
\end{equation}
described by the decomposition
of $\tau$ into $SU(3)$ irreducible representations:
\begin{equation}
(1+1)+(8+8)+(6+\bar{6})+(3+\bar{3})+(3+\bar 3)\,.
\end{equation}
We note that ${\cal W}_1, \ldots , {\cal W}_{4}$ is the space determining the 
torsion of the
$U(3)$ structure \cite{GrayHervella}.

A closer analysis of these classes and of how they are determined will 
let us determine many important properties of the manifold.
The five ${\cal W}_{i}$ are expressed in terms of $d J$ and $d \Psi$ 
in the following manner \cite{chiossi}:
\begin{equation}
\begin{array}{rclcrcl}
{\cal W}_1 & \leftrightarrow & \left[ dJ\right]^{(3,0)}\,,  & \quad &
{\cal W}_2 & \leftrightarrow & \left[ d\Psi\right]_0^{(2,2)}\,,  \\
{\cal W}_3 & \leftrightarrow & \left[ dJ\right]_0^{(2,1)}\,,  & \quad &
{\cal W}_4 & \leftrightarrow & J \wedge dJ\,,  \\
{\cal W}_5 & \leftrightarrow & \left[ d\Psi\right]^{(3,1)}\,. & &&&
\end{array}
\label{classes}
\end{equation}
The subscript $0$ is used to remove all the forms which are not 
primitives, i.e. $\beta \in \Lambda_{0}^{(2,2)}$ if $J \wedge \beta = 
0$, and $\gamma \in \Lambda_{0}^{(2,1)}$ if $J \wedge \gamma =0$.
A more precise definition \cite{chiossi} comes as follows.
The first class ${\cal W}_1$ is given by two real numbers ${\cal W}_1 =
{\cal W}_1^+ + {\cal W}_1^-$.
These components are selected by taking the exterior product with 
$\Psi$ and, using (\ref{psipm}), one can define
\begin{equation}
\begin{array}{rcccl}
d \psi_+ \wedge J &=& \psi_+ \wedge dJ &=& {\cal W}_{1}^+\, 
J\wedge J\wedge J\,,\\
d \psi_- \wedge J &=& \psi_- \wedge dJ &=& {\cal W}_{1}^-\, J\wedge J\wedge 
J\,,
\end{array}
\label{W1}
\end{equation}
where $J^3$ is essentially the volume form.
Similarly, one can see that the $\Lambda^{(2,2)}_{0}$ part of $d\Psi$ 
is determined by two $(1,1)$--forms, since one removes all the 
non--primitive forms.
The two components of ${\cal W}_2$ (${\cal W}_2 =
{\cal W}_2^+ + {\cal W}_2^-$) are defined as follows,
\begin{equation}
\begin{array}{rcl}
\left( d \psi_+ \right)^{2,2} &=&  {\cal W}_{1}^+ \,J\wedge J+ 
{\cal W}_2^+\, \wedge J\,,\\
\left( d\psi_-  \right)^{2,2}&=&  {\cal W}_{1}^- \,J\wedge J+ 
{\cal W}_2^-\, \wedge J\,.
\end{array}
\label{W2}
\end{equation}
To define the components of ${\cal W}_4$ and ${\cal W}_5$, one further 
needs the introduction of the contraction operator
\begin{equation}
\begin{array}{rcl}
\lrcorner \, : \: \bigwedge{}^k \, T^* \otimes \bigwedge{}^n \, T^* &\to& 
\bigwedge{}^{n-k} \, T^* \\[3mm]
\left( L_{(k)} , M_{(n)} \right) &\mapsto& \displaystyle 
\frac{1}{n!} \, \left(\begin{array}{c} n \\ k 
\end{array} \right) \, L^{a_1 \ldots a_k} M_{a_1 \ldots a_n} 
e^{a_{k+1}} \ldots e^{a_n}\,
\end{array}
\label{contraction}
\end{equation}
with the convention that $e^{1}Ê\wedge e^2 \,\lrcorner\, e^1 \wedge 
e^2 \wedge e^3 \wedge e^4 = e^3 \wedge e^4$.
Using this operator, we can now define ${\cal W}_4$ and ${\cal W}_5$, 
which are the 
following one--forms:
\begin{eqnarray}
{\cal W}_4 &=& \frac12 \, J \, \lrcorner \, dJ \,, \label{W4}\\
{\cal W}_5 &=& \frac12 \, \psi_+ \lrcorner \, d\psi_+ \,. \label{W5}
\end{eqnarray}
${\cal W}_3$ can be read off from
\begin{equation}
dJ^{(2,1) } = \left[
J \wedge {\cal W}_4\right]^{(2,1)} + {\cal W}_3 
\;.
\label{defw3}
\end{equation}

Examples of manifolds which fall into this classification are given by 
Calabi--Yau spaces.
In this case one has strict $SU(3)$ holonomy, namely the torsion is zero
and therefore the r.h.s of these equations vanishes.
This is equivalent to the fact that we have a K\"ahler structure 
satisfying $dJ = 0$ and a closed holomorphic 3--form of constant norm $d\Psi=0$.
However, we will see that there are several other cases of physical 
interest.

Depending on the classes of torsion one can obtain many different 
types of manifolds (complex or not), whose properties have often been 
studied in the mathematical literature. 
Here, we will list \cite{GrayHervella}
those manifolds which we will be using in the following.
First of all it is important to distinguish between complex and 
non--complex manifolds.

\vskip0.2cm
\noindent
{\it i) Complex manifolds}

\vskip0.2cm

The components of the Nijenhuis tensor 
completely determine ${\cal W}_1 \oplus {\cal W}_{2}$, therefore:
\begin{equation}
{\cal W}_1  = {\cal W}_{2} = 0 \Leftrightarrow \hbox{Hermitian manifold.} 
\label{hermitiancond}
\end{equation}
These equations tell us that these manifolds have a complex 
structure $J$ such that the Hodge type is preserved upon 
differentiation, 
\begin{equation}
d J \in \Lambda^{(2,1)} \oplus \Lambda^{(1,2)} \quad \hbox{and} \quad 
d\Psi \in \Lambda^{(3,1)} \oplus \Lambda^{(1,3)}\,.
\label{hermitian}
\end{equation} 
The condition (\ref{hermitiancond}) is also equivalent to
\begin{equation}
d \alpha^{(0,2)} = 0\,, \qquad \forall \; \alpha \in \Lambda^{(1,0)}\,.
\label{da02}
\end{equation}
Inside the class of {\it complex manifolds} we find
\begin{equation}
\begin{array}{rcl}
{\cal W}_1={\cal W}_2={\cal W}_4=0\,, \quad
\tau \in {\cal W}_{3} \oplus {\cal W}_{5}
& \Leftrightarrow & \hbox{balanced
manifolds,}  \\[3mm]
{\cal W}_1={\cal W}_2={\cal W}_4={\cal W}_5=0\,, \quad
\tau \in {\cal W}_{3} & \Leftrightarrow & \hbox{special--hermitian
manifolds,}  \\[3mm]
{\cal W}_{1}={\cal W}_{2}={\cal W}_{3}={\cal W}_{4}=0\,, \quad
\tau \in {\cal W}_5 & \Leftrightarrow & 
\hbox{K\"ahler manifolds,} \\[3mm]
{\cal W}_1=
{\cal W}_2={\cal W}_3={\cal W}_4=
{\cal W}_5=0\,,\quad
\tau =0 & \Leftrightarrow & \hbox{Calabi--Yau manifolds.}
\end{array}
\label{complex}
\end{equation}
Another class of complex manifolds which does not lie in a definite 
sector of the above classification is given by the {\it 
Strong K\"ahler Torsion} (SKT) manifolds.
These manifolds are such that 
\begin{equation}
 \partial \bar{\partial} J = 0\,, \quad \hbox{with } d J \neq 0\,.
\label{SKT}
\end{equation}
We will see that these manifolds are related to solutions of the 
Bianchi identity $dH = 0$.
A result of \cite{strongKT} tells us that if a 
manifold of dimension $d\geq 6$ 
is 
special--hermitian (i.e. $\tau \in {\cal W}_{3}$) then it {\it cannot} be SKT.
A short alternative proof  is the following: a complex manifold with 
${\cal W}_{4}=0$ satisfies $J \wedge d J = 0$.
If we now take a $\bar \partial$ derivative on it we get
\begin{equation}
\bar\partial (J\wedge d J )=
\bar\partial J \wedge \partial J + J \wedge \bar \partial \partial J=0\,,
\label{mah}
\end{equation}
where the last term vanishes because of the SKT condition.
What remains can also be written as $\overline{\partial J} \wedge \partial 
J =0$ from which it follows that the manifold has also to be 
K\"ahler $\partial J = \bar \partial J = 0$.

\vskip0.2cm
\noindent
{\it ii) Non--complex manifolds}
\vskip0.2cm

Some {\it non--complex manifolds} which will occur later on are 
\begin{eqnarray}
\tau \in {\cal W}_1  &\Leftrightarrow& \hbox{nearly--K\"ahler manifolds,} 
\label{nearlyKaehler}
\\
\tau \in {\cal W}_2  &\Leftrightarrow& \hbox{almost--K\"ahler manifolds.} 
\label{almostKaehler}
\end{eqnarray}
Again, in terms of forms, nearly--K\"ahler manifolds are defined by a 
closed $(3,0)$--form and an almost--complex structure satisfying
\begin{equation}
d J \in \Lambda^{(3,0)}\,, \quad d\Psi = {\cal W}_1 \, J \wedge J\,.
\end{equation}
Almost--K\"ahler manifolds are instead defined by a closed 
almost--complex structure, but a $(3,0)$ form satisfying
\begin{equation}
d J = 0\,, \quad d \Psi \in \Lambda^{(2,2)}_{0}\,.
\end{equation}

The last definition we want to introduce here and that will be very 
important in our follow--up is given by the 
so--called {\it half--flat} manifolds \cite{chiossi}.
These manifolds can be either complex or not complex and satisfy 
\begin{equation}
\tau \in {\cal W}_1^- \oplus {\cal W}_2^- \oplus {\cal W}_3\,,
\label{halfflat}
\end{equation}
which implies
\begin{equation}
d \psi_{+} = 0 \quad \hbox{and} \quad J\wedge dJ = 0\,.
\end{equation}
We will see that such manifolds will be useful in constructing 
seven--dimensional manifolds of $G_{2}$ holonomy.

%
%
%

Finally, we want to mention a result of \cite{GrayHervella} about 
conformally rescaled metrics.
Conformally rescaling the metric adds 
components to the torsion in ${\cal W}_4 \oplus {\cal W}_5$.
In particular, the following combination
remains fixed \cite{chiossi},
\begin{equation}
3 {\cal W}_4 + 2 {\cal W}_5 \;. 
\label{comb}
\end{equation}
For instance, 
conformally rescaled Calabi--Yau manifolds will be the proper subset of 
the manifolds in the class 
\begin{equation}
\tau \in {\cal W}_4 \oplus {\cal W}_5
\label{confCY}
\end{equation}
for which
$2 {\cal W}_4 =- 3 {\cal W}_5$.

Let us now briefly review a construction by Hitchin \cite{Hitchin} whereby
a six--dimensional half--flat manifold is lifted
to a seven--dimensional manifold with $G_2$ holonomy.
This is achieved as follows.  Starting with a six--dimensional
manifold $M$ equipped with an $SU(3)$ structure $(M,J,\Psi)$,
a seven--dimensional manifold is then constructed as a warped
product $X_7= M \times I$, where $I\subset {\mathbb R}$, and the $G_{2}$ 
structure is defined by
\begin{equation}
\phi =  J\wedge dr + \psi_{+}\,,
\label{G2structure}
\end{equation}
where $dr$ is the line element on $I$ and
$(J,\psi_+,\psi_-)$ are now $r$--dependent.
The $G_2$ structure of $X_7$ satisfies
\begin{eqnarray}
d\phi&=& \left(\hat dJ-{\partial\psi_+\over\partial r}
\right)\wedge dr+\hat d\psi_+\, ,\nonumber\\
d\star\phi&=& \left(\hat d\psi_-+J\wedge{\partial J\over\partial 
r}\right)\wedge dr+J\wedge\hat dJ\, ,\label{intervall}
\end{eqnarray}
where the forms have now been promoted to seven--dimensional forms 
and $\hat{d}$ denotes the six--dimensional differential operator.
Since $M$ is half--flat, its $SU(3)$ structure is such that 
$\hat d \psi_+ = J \wedge \hat d J = 0$.  
Therefore, demanding that the seven--dimensional manifold $X_7$ has
$G_2$ holonomy, i.e. $d \phi = d \star \phi =0$, yields
\begin{equation}
\left\{ 
\begin{array}{rcl} 
\hat{d} J &=& \displaystyle \frac{\partial \psi_{+}}{\partial r}\,, \\  
\hat{d} \psi_{-} &=& \displaystyle- J \wedge \frac{\partial J}{\partial r}\,.
\end{array}
\right.
\label{hitchin}
\end{equation}
These are the flow equations of Hitchin \cite{Hitchin}.

\section{${\cal N}=1$ supersymmetric compactifications of 
heterotic string theory}

In the following, we will be interested in ${\cal N}=1$ supersymmetric
compactifications of heterotic string theory on spaces with metric
given by
\begin{equation}
ds^2 = g^0_{MN} \, dx^M \otimes dx^N = 
{\rm e}^{2\Delta(y)} \left( dx^\mu \otimes dx^\nu \, \hat g_{\mu\nu}(x)
+ dy^m \otimes dy^n \, \hat g_{mn}(y)\right)\,.
\label{metric}
\end{equation}
Here ${\hat g}_{\mu\nu}(x)$ denotes the metric of a four--dimensional
maximally symmetric spacetime and 
$\Delta$ denotes a warp factor which we take to only depend on the
internal coordinates $y^m$.  
It can be shown \cite{Strominger:1986uh,deWit:1987xg} that the
ten--dimensional supersymmetry equations (in the absence of gaugino
condensates) can be cast into the following form
\begin{subequations}
\begin{eqnarray}
\delta \psi_{M} &=& {\cal D}_M \epsilon  \equiv
\nabla_{M}\epsilon - \frac14 H_{M} \epsilon \,, 
\label{gravitinor}\\
\delta \chi &=& - \frac14 \,\Gamma^{MN}\epsilon 
\,F_{MN}\,,
\label{gauginor}\\
\delta \lambda &=&  \nabla\!\!\!\!\slash \phi + \frac{1}{24} \, 
H\, \epsilon  \,,
\label{dilatinor}
\end{eqnarray}
\label{susyr}
\end{subequations}

\noindent
where $H \equiv \Gamma^{MNP} H_{MNP}$, $H_{M} 
\equiv H_{MNP} \,\Gamma^{NP}$, 
and where the covariant derivative $\nabla$ is 
constructed from the rescaled metric $g_{MN} = {\rm e}^{-2 \phi} g_{MN}^0$.
Necessary and sufficient conditions for 
${\cal N}=1$ spacetime supersymmetry in four dimensions were
derived in \cite{Strominger:1986uh} and are given by:
\begin{enumerate}
\item the four--dimensional spacetime has to be Minkowski, i.e. $
{\hat g}_{\mu \nu} = \eta_{\mu \nu}$;
\item the internal six--dimensional manifold has to be complex, i.e.
the Nijenhuis tensor 
$N_{mnp} $ has to vanish;
\item up to a constant factor, 
there is exactly one holomorphic $(3,0)$--form $\omega$, 
whose norm is related to 
the complex structure $J$ by
\begin{equation}
\star d \star 
J = 
i  \left( \bar \partial - \partial\right)\, \log ||\omega|| \,;
\label{ddagger}
\end{equation}
\item the Yang--Mills background field strength must be a $(1,1)$-- form
and must satisfy
\begin{equation}
\hbox{tr} F \wedge F = \hbox{tr} \, \tilde R \wedge 
\tilde R  - i \, \partial \bar \partial J \,
\label{ddJ}
\end{equation}
as well as 
\begin{equation}
\label{FJ}
F_{mn}J^{mn} = 0\,;
\end{equation}
\item the warp factor $\Delta$ and the dilaton $\phi$ are determined by
\begin{eqnarray}
\Delta (y) &=& \phi (y) + {\rm constant} \;,\nonumber\\
\phi ( y) & =& \frac{1}{8} \log ||\omega|| + {\rm constant} \;;
\end{eqnarray}
\item
the background three--form $H$ is determined in terms of $J$ by
\begin{equation}
H = \frac{i}{2} \,\left(\bar\partial - \partial\right) J\,,
\label{HJ}
\end{equation}
where $i (\partial - { \bar \partial} ) = dx^n \, J_n\,^m \,
\partial_m$. 

\end{enumerate}
Inspection of (\ref{ddJ}) shows that if 
${\rm tr} 
\,{\tilde R} \wedge {\tilde R}$ is non--vanishing, then it has to be a
$(2,2)$--form for consistency.

We note that the integrability associated with the vanishing of the internal
gravitino equation (\ref{gravitinor}) (cf. appendix B),
\begin{equation}
\left[{\cal D}_m, {\cal D}_n\right] \eta_+ = \frac14  
\tilde{R}_{mn}{}^{pq} \Gamma_{pq} \eta_+ = 0 \,,
\label{integr}
\end{equation}
implies the vanishing of
the Ricci two--form 
\begin{equation}
\tilde{R}_{mn}{}^{pq} J_{pq} = 0\,,
\label{Ricciform}
\end{equation}
where $\tilde{R}_{mn}{}^{pq}$ denotes the generalized Riemann tensor
constructed out of 
\begin{equation}
{\cal D}_{m} \equiv \partial_{m} + \frac14 \left({\omega_{m}}^{np} - 
{H_m}^{np}\right) \Gamma_{np}\,.
\label{covd}
\end{equation}
Here $H_m\,^{np}$ denotes a torsion term. 
We also note that in general it does {\it not} follow that the 
generalized Ricci tensor ${\tilde R}_{mn} =\tilde{R}^q\,_{mqn}$
vanishes.  
Usually, the vanishing of the Ricci tensor is derived by
multiplying (\ref{integr}) with $\Gamma^n$ and using that $R_{m[npq]} =0$.
In general, however, ${\tilde R}_{m[npq]} \neq 0$ because of the torsion
terms, and hence one cannot conclude that ${\tilde R}_{mn} =0$.

Next we will reformulate the conditions just mentioned 
in terms of torsional constraints, using the language of section 
\ref{mathpart}.  
Since the internal manifold is taken to be complex,
it immediately follows from (\ref{hermitiancond}) that
\begin{equation}
{\cal W}_1 = {\cal W}_2 = 0\,.
\label{cW}
\end{equation}
The torsion is therefore left in 
\begin{equation}
\tau \in {\cal W}_3 \oplus {\cal W}_4 \oplus {\cal W}_5 \,,
\label{torsion1}
\end{equation}
but it cannot be completely generic, because there is one 
further geometric constraint to be satisfied, namely (\ref{ddagger}).
This equation relates the dual of the complex structure to the 
holomorphic (3,0)--form and therefore can be interpreted as a relation 
among the ${\cal W}_4$ and ${\cal W}_5$ classes.
The ${\cal W}_4$ class is determined by $J \wedge d J$ which, using 
the duality relation $\star J = \frac12 J \wedge J$, 
can be interpreted as $d \star J$.
This implies that information about this class is encoded in the 
left--hand side of equation (\ref{ddagger}), as this is given by the 
one--form $\star d \star J$.
Moreover, from the definition of ${\cal W}_4$ (\ref{W4}), it follows that it 
must be described by a one--form, so it is interesting to establish the 
precise relation among the two quantities.
Following the definition of the contraction operator 
(\ref{contraction}), we can rephrase the Hodge star dual and show that
\begin{equation}
{\cal W}_4 =\frac12 \, J \, \lrcorner \, dJ = 
\frac12\, J \cdot \left( \star d \star J\right)\,.
\label{W4Stro}
\end{equation}
The proof follows directly from the definition of the contraction operator 
\begin{equation}
 J \, \lrcorner \, dJ = \frac32 \; 
J^{sn} \; dx^p \;\nabla_{[s} J_{np]} = dx^n\; {J_s}^p\, \nabla_p 
 {J_n}^s \,,
\label{proof1}
\end{equation}
and of the Hodge dual:
\begin{equation}
 J \cdot \left( \star d \star J\right) = - dx^n\; {J_n}^s\; \nabla_p 
 {J_s}^p =  dx^n\; {J_s}^p\; \nabla_p 
 {J_n}^s \,.
\label{proof2}
\end{equation}
Going back to (\ref{ddagger}), and in order to determine the precise relation 
between ${\cal W}_4$ and ${\cal W}_5$, 
we better consider multiplying (\ref{ddagger})
with $J$.
In this way the equation gets simplified to 
\begin{equation}
{\cal W}_4 =-\frac12\, d \log ||\omega||\,,
\label{deq}
\end{equation}
which gives a further constraint on ${\cal W}_4$, namely that it 
is an exact real  
1--form.
On the right hand side of this equation we find the norm of the holomorphic 
form, which is related to ${\cal W}_5$.
Our classification of the torsion relies on the definition of a 
unit norm (3,0)--form, which in this case is simply
\begin{equation}
\Psi = \frac{\omega}{||\omega||} \,.
\label{Psidef}
\end{equation}
This form is not holomorphic anymore for a generic dilaton profile 
(which is then related to the $\omega$ norm) and that implies 
${\cal W}_5 \neq 0$.
From the definition of the unit--norm (3,0)--form (\ref{Psidef}) it 
follows that 
\begin{equation}
d \psi_+ = \frac12 \, \left( d \Psi + d \bar \Psi\right) = - d \log ||\omega|| 
\wedge \psi_{+}\,.
\label{dpsi}
\end{equation}
The contraction with $\psi_+$ will therefore lead to
\begin{equation}
{\cal W}_5 = \frac12 \psi_+ \, \lrcorner \, d \psi_+ = d \log ||\omega||\,,
\label{W5Stro}
\end{equation}
and this finally translates into 
\begin{equation}
2 \,{\cal W}_4 + {\cal W}_5 =0\,.
\label{Wrelation}
\end{equation}
Since the remaining equations to obtain supersymmetric 
solutions do not give rise 
to further geometrical constraints, but rather to equations in the gauge 
sector, we can conclude that the torsion of the complex manifolds 
giving rise to supersymmetric compactifications of heterotic string
theory has to satisfy
\begin{equation}
\begin{array}{rcl}
\tau &\in& {\cal W}_3 \oplus {\cal W}_4 \oplus {\cal W}_5\,, \\[3mm]
\hbox{ with } && 2\, {\cal W}_4 + {\cal W}_5 = 0 \\[3mm]
\hbox{ and } && {\cal W}_4\,, \,{\cal W}_5 \quad
\hbox{ exact and real.}
\end{array}
\label{torsion}
\end{equation}
Note that the condition of 
${\cal W}_4$ and of ${\cal W}_5$ being exact is a global condition.
Conversely, (\ref{torsion}) implies that the six--dimensional manifold is
complex, that (\ref{ddagger}) holds and that there exists a holomorphic
$(3,0)$--form without zeroes or poles, 
as we will now show.  Let ${\cal W}_5= d\Sigma$ 
with $\Sigma$ real and
define $\omega = {\rm e}^{\Sigma} \Psi$.  One has $d \psi_+ = \psi_+ \wedge 
d \Sigma$
and hence \cite{chiossi} $d \psi_{-} = \psi_+ \wedge J \cdot d \Sigma$.
Since $\psi_+ \wedge J \cdot d \Sigma = \psi_- \wedge d \Sigma$, 
it follows that
$d \Psi = \Psi \wedge d \Sigma$, and hence $d \omega =0$.

We can now try to discuss which classes of the manifolds discussed in section 
\ref{mathpart} can be used as consistent solutions of the torsion 
constraints (\ref{torsion}).

The easiest way to solve (\ref{torsion}) non--trivially is to 
consider manifolds which have both ${\cal W}_4$ and ${\cal W}_5$ vanishing 
independently.
This implies that the torsion is left only in the ${\cal W}_3$ sector, 
which means that a particular class of solutions is 
given by {\it special--hermitian manifolds}.
In this case the solutions will have a constant dilaton profile and 
therefore a closed (holomorphic) (3,0)--form of constant norm.
The only difference between this new class of solutions and the known 
Calabi--Yau manifolds is given by a non--trivial 
three--form $H$, which drives the change of the geometry and makes the 
manifold non--K\"ahler.  
Examples of special--hermitian manifolds are the nilmanifolds
\cite{salamon} and the Moishezon manifolds \cite{Gutowski:2002bc}.

More general classes of solutions have a non--constant 
dilaton profile, as the torsion in the ${\cal W}_4$ and ${\cal 
W}_5$ classes is not vanishing anylonger.
However, as shown in (\ref{torsion}), it is not sufficient to find a
generic manifold lying in these classes, because the one--forms
defining ${\cal W}_4$ and ${\cal W}_5$ have to be exact and linearly
dependent.
It is interesting to point out that the definition of ${\cal W}_4$ in terms 
of $\theta \equiv J \cdot \left(\star d \star J\right)$, as follows from 
(\ref{W4Stro}), can be translated as a condition on the so--called 
Lee--form, which is precisely given by $\theta$ \cite{Lee}.
This form defines the {\it balanced} manifolds, which are those 
with vanishing $\theta$, and the {\it conformally balanced} 
manifolds, which have $\theta$ exact, i.e. $\theta = d \Lambda$.
The conformally balanced manifolds are also known to admit a 
holomorphic (3,0)--form \cite{Papadopoulos:2002gy}, 
and therefore could be used as proper solutions 
of the torsional constraints.
Of course, in this class one has to select those manifolds whose 
${\cal W}_4$ and ${\cal W}_5$ classes are linearly dependent through 
the relation (\ref{Wrelation}).

An interesting subcase is given by the manifolds which lead to a closed 
three--form flux $dH = 0$, which are the so--called SKT 
manifolds (\ref{SKT}).
In this case one may solve the gauge sector by the 
usual ``spin in the gauge" procedure.
However it has been proved \cite{Ivanov:2000fg,Papadopoulos:2002gy,
Gauntlett:2001ur,Gauntlett:2002sc} 
that a compact, conformally balanced, SKT manifolds admits holonomy 
in $SU(3)$ only if it is also K\"ahler (and then CY).
An example of a SKT manifold is the  
$S^3 \times S^3$ manifold, which for a 
proper choice of the almost--complex structure may also be shown to be 
complex \cite{Gutowski:2002bc}.  
However it does not admit a holomorphic $(3,0)$--form and therefore it
is not conformally balanced.

We finish this section with some comments on the use of manifolds derived 
by rescalings of the metric.
In section \ref{mathpart} we mentioned that a class of manifolds with torsion 
in ${\cal W}_4$ and ${\cal W}_5$ is given by conformally rescaled 
metrics, among which are the conformally rescaled Calabi--Yau manifolds.
These latter manifolds {\it cannot} be used as solutions because the torsion 
has to satisfy (\ref{Wrelation}),
whereas the 
conformally rescaled CY metrics have to fulfill
$3{\cal W}_4 + 2 {\cal W}_5 =0$. 
 
Nevertheless, the trick of conformally rescaling the metric may yield more
general solutions to the torsional constraints.
Indeed, if one finds a complex manifold such that the ${\cal W}_4$ 
and ${\cal W}_5$ classes are defined by exact real forms,
\begin{equation}
{\cal W}_4 = d \Lambda\,, \quad {\cal W}_5 = d \Sigma\,,
\label{W4W5resc}
\end{equation}
then their linear combination is also an exact form, and in particular
\begin{equation}
2 {\cal W}_4 +{\cal W}_5 = d \left(2\Lambda+ \Sigma\right) \equiv d \,f\,.
\label{df}
\end{equation}
From this equation it follows that one can now satisfy the torsional 
constraints (\ref{torsion}) upon rescaling the metric by
\begin{equation}
g \to {\rm e}^{2 f} g\,,
\label{rescal}
\end{equation}
as follows.  
Under the rescaling (\ref{rescal}) it can be shown \cite{chiossi}
that 
\begin{eqnarray}
{\cal W}_4 & \to& {\cal W}_4 + 2 d f \;,\nonumber\\
{\cal W}_5 & \rightarrow& {\cal W}_5 - 3 d f \;,
\end{eqnarray}
and hence
\begin{equation}
2 {\cal W}_4 + {\cal W}_5 - df =0 \to 2 {\cal W}_4 + {\cal W}_5 =0 \;.
\end{equation}
The holomorphic $(3,0)$--form associated with (\ref{df}) is given by
$\omega = {\rm e}^{\Sigma} \Psi$.  
Upon rescaling of the metric, $\omega$
will rescale as $\omega \to {\rm e}^{3f} \omega \equiv 
{\hat \omega}$.  
Note, however, that ${\hat \omega}$ is not anylonger
holomorphic.  
A $(3,0)$--form, which is also holomorphic after the
rescaling, is instead given by $\omega$, whose norm in the rescaled
metric is given by $||\omega|| = {\rm e}^{\Sigma - 3 f}$.

\section{The heterotic string on the Iwasawa manifold}
\label{Iwasawa}

In this section we will construct an explicit example of a manifold in 
the class described by (\ref{torsion}) which solves the supersymmetry equations
(\ref{susyr}).
We will, for simplicity, consider 
{\it special--hermitian manifolds}, 
which are manifolds whose torsion is contained 
only in ${\cal W}_{3}$.
These manifolds are the only complex manifolds which are also half--flat,
the latter being upliftable to $G_{2}$ spaces.
Since for such manifolds ${\cal W}_4 = {\cal W}_5 = 0$, the dilaton stays 
constant and equation (\ref{ddagger}) is identically solved, 
because both terms are zero independently.

To further simplify the analysis we will consider the so--called 
six--dimensional nilmanifolds.
These manifolds are constructed from simply--connected nilpotent Lie 
groups $G$ by quotienting with a discrete subgroup $\Gamma$ of $G$ for 
which $G\backslash \Gamma$ is compact\footnote{Such $\Gamma$ subgroup exists 
if the structure equations of its Lie algebra are rational 
\cite{Malcev,Nomizu}.}.
There are 34 classes of such manifolds and they do not admit a K\"ahler 
metric, implying that any solution involving these manifolds will yield 
a non--trivial  departure from the well known Calabi--Yau examples.

Out of these 34 classes, 
18 admit a complex structure \cite{AGS,salamon}.
An interesting example coming from a two--step algebra\footnote{An 
algebra $\mathfrak g$ is $n$--step nilpotent if ${\mathfrak g}^n  
= 0$.} inside this 
classification is given by the Iwasawa manifold, which is going to be 
the one we will analyze in the following.
It is interesting to point out that for any of the 18 classes one
can choose a complex structure which is compatible with the metric and 
which has 
${\cal W}_4= 0$ \cite{strongKT}.  
Moreover, since on a nilmanifold which admits a complex structure one
can choose a basis of one--forms such that $d \alpha \in
\Lambda^{(2,0)}$ for any given $\alpha \in \Lambda^{(1,0)}$
\cite{salamon}, also ${\cal W}_5=0$.  
One can check that one can choose a complex structure such that both
conditions can be satisfied simultaneously and that therefore the
resulting torsion lies entirely in ${\cal W}_3$.

The Iwasawa manifold is a nilmanifold obtained as the compact 
quotient space $M = \Gamma \backslash G$, where $G$ is the complex 
Heisenberg group and $\Gamma$ is the subgroup of the Gaussian integers.
The complex Heisenberg group is given by the set of matrices
\begin{equation}
G = \left\{  \left( \begin{tabular}{ccc}1 & $z$ & $u$ \\ 0 & 1 & $v$ \\ 0 & 
0 & 1\end{tabular}\right) \; : \; u,v,z \in {\mathbb C} \right\}
\label{heisenberg}
\end{equation}
under multiplication and the discrete subgroup $\Gamma$ is defined by 
restricting $u$, $v$, $z$ to Gaussian integers:
\begin{equation}\left\{
\begin{array}{rcl}
v & \to & v+ m\,,  \\
z & \to & z + n\,,  \\
u & \to & u + p +nv\,,   
\end{array}
\right.
\quad m,n,p\in {\mathbb Z} \oplus i \, {\mathbb Z}\,.
\label{Gauss}
\end{equation}

Left invariant one--forms on $G$ are given by
\begin{equation}
\begin{array}{rcl}
\alpha &=& dz\,,\\
\beta & = & dv\,,  \\
\gamma & = & -du + z \, dv\,.  \\ 
\end{array}
\label{leftforms}
\end{equation}
These are also holomorphic $(1,0)$--forms with respect to the standard 
complex structure as we will see later.
The Lie bracket is determined by the equations
\begin{equation}
 \begin{array}{rcl}
d \alpha & = & 0\,,  \\
d \beta &=& 0\,,\\
d\gamma & = & \alpha \wedge \beta\,.
\end{array} 
\label{dal}
\end{equation}
Setting 
\begin{equation}
\alpha = e^1 + i \, e^2\,, \quad \beta = e^3 + i \, e^4\,, \quad 
\gamma = e^5 + i\, e^6\,,
\label{real}
\end{equation}
one can introduce a real basis for the dual of the Lie algebra 
corresponding to $G$, which is also the cotangent space.
In this way the real form of equation (\ref{dal})  is given by
\begin{eqnarray}
d e^i &=& 0 \;\;\;,\;\;\; i = 1, \dots ,4 \;, \nonumber\\
de^5 &=& e^1\wedge e^3 + e^4 \wedge e^2 \;\;\;, \nonumber\\
de^6 &=& e^1\wedge e^4 + e^2 \wedge e^3 \;\;\;.
\end{eqnarray}
Thus we see that, since there is a 
four--dimensional kernel of 
the differential map, one can find a 
geometrical description of the 
manifold $M$ as a principal ${\mathbb 
T}^2$--fibration over ${\mathbb T}^2 \times {\mathbb T}^2 $.  We also
note that there is no section of the fibration, so the base does not lie
embedded in $M$.

An important and established mathematical result \cite{AGS,ketse} is that the 
set of almost complex structures on $M$ which are compatible with the 
Riemannian metric $ds^2 = \sum_{i=1}^6 (e^i)^2$ 
are isomorphic to ${\mathbb C \mathbb P}^3$.
In particular one can define the fundamental 2--form at an arbitrary 
point of the moduli space of almost complex structures as 
\begin{equation}
J = e^5 \wedge \left( \cos \theta \,e^6 + \sin \theta \, f^1 \right) - 
f^2 \wedge \left(\cos \theta \, f^1  - \sin \theta \,e^6\right) + f^3 
\wedge f^4\,,
\label{Jiwasawa}
\end{equation}
where $f^i$ are an $SO(4)$ rotated basis for $e^1$, \ldots, $e^4$, with 
the definition $f^i = {P^i}_j e^j$, $P\in SO(4)$.  
Note that inequivalent complex structures arise only for $P \in SO(4)/U(2)$.
The appropriate $(1,0)$--form basis relative to the almost complex 
structure (\ref{Jiwasawa}) is given by
\begin{equation}
\begin{array}{rcl}
\alpha & = & \cos \theta \, f^1  - \sin \theta \,e^6 + i \, f^2\,,  \\
\beta & = & f^3 + i \, f^4\,,  \\ 
\gamma & = & e^5 + i \, \left(\cos \theta\, e^6 +  \sin \theta \, f^1 
\right) \,.
\end{array}
\label{10forms}
\end{equation}
In terms of these $(1,0)$--forms 
the almost complex structure is given by
\begin{equation} 
J = \frac{i}{2}\,\left(\alpha \wedge 
\bar \alpha + \beta\wedge \bar \beta+ \gamma \wedge \bar \gamma\right)\,,
\end{equation}
and the norm one $(3,0)$--form by
\begin{equation}
\Psi = \alpha \wedge \beta \wedge \gamma\,.
\label{30form}
\end{equation}

If we want the Iwasawa manifold to be a solution of the torsional
constraints, we must choose the parameters in (\ref{Jiwasawa}) such
that the torsion lies in ${\cal W}_3 \oplus {\cal W}_4 \oplus {\cal
W}_5$, satisfying (\ref{torsion}).  
Demanding the torsion to be in these classes, it has been shown in
\cite{AGS} that ${\cal W}_4=0$.

The moduli space of complex structures turns out to have two disconnected
components which are given by
\begin{equation}
\theta = 0 \hbox{  and  } P = I\,, \quad 
\label{point}
\end{equation}
and
\begin{equation}
\quad \theta = \pi \hbox{  and  } P \in SO(4)\,.
\label{line}
\end{equation}
The case $\theta =0$ and $P$ the identity gives the standard complex 
structure $J_0$, whereas (\ref{line}) describes a complex projective line
of inequivalent complex structures
(an `edge').
We will show below that ${\cal W}_5=0$ on both components
of the moduli space.

\subsection{The standard complex structure $J_0$}

Consider picking the standard complex structure $J_0 = e^1\wedge e^2
+e^3\wedge e^4 + e^5\wedge e^6$.  
For this choice of $J$ the manifold is complex and therefore we 
already satisfy the first geometrical constraint.
The $(1,0)$--forms are then given by
\begin{equation}
\alpha =   e^1 + i e^2 \,, \quad
\beta =e^3 + i e^4 \,, \quad
\gamma = e^5 + i e^6\,.
\end{equation}
Complex coordinates $(z,v,u)$ can be introduced accordingly, in order 
to evaluate the various geometrical quantities explicitly:
\begin{equation}
\alpha = dz \,, \quad
\beta = dv\,, \quad
\gamma = - du + z dv  \,.
\end{equation}
These are the natural coordinates inherited from the definition of the
manifold as a coset (c.f (\ref{leftforms})).
The forms $\alpha, \beta$ and $\gamma$ are holomorphic
with respect to $J_0$.  
The differentials $dz, dv$ and $du$ are closed differentials, i.e.
$d^2 z = d^2 v = d^2 u =0$.  
This is consistent with evaluating
$d \gamma = e^1\wedge e^3 + e^4\wedge e^2 + 
i (e^1\wedge e^4 + e^2\wedge e^3) = dz \wedge dv$ (cf. (\ref{dal})).
The complex structure in these coordinates is given by
\begin{equation}
J_0 = 
\frac{i}{2} \left[ dz \wedge d\bar{z} + dv\wedge d {\bar v} + (du - z dv)\wedge
(d{\bar u} - {\bar z}  d{\bar v}) \right] \;,
\end{equation}
and the metric is 
\begin{equation}
ds^2 =  \left[dz d\bar{z} + dv d {\bar v} + (du - z dv)
(d{\bar u} - {\bar z}  d{\bar v})\right]\,.
\label{metricIwa}
\end{equation}
This makes it evident that the Iwasawa manifold can be viewed as
a ${\mathbb T}^2$ 
fibration over a ${\mathbb T}^2 \times {\mathbb T}^2$
base.

We find the following closed forms modulo exact forms
(note that the Euler number of the torus fibration is zero)
\begin{equation}
\begin{array}{cccc}
\underline{b_1=4:} &
 \al \; , \; \be  & &      \bar{\al}\; , \;
\bar{\be}\nonumber\\[2mm]
\underline{b_2=8:}&\al\ga \; , \; \be\ga &
\al\albar \; , \; \al\bebar \; , \; \be\albar \; , \; \be\bebar
                &
\albar\gabar \; , \; \bebar\gabar\nonumber\\[2mm]
\underline{b_3=10:} & \al\be\ga &
\al\ga\albar \; , \; \al\ga\bebar \; , \; \be\ga\albar \; , \;
\be\ga\bebar \;\;\;\;\;\;\;\;
\al\albar\gabar \; , \; \be\albar\gabar \; , \; \al\bebar\gabar \; ,
\; \be\bebar\gabar  & \albar\bebar\gabar\nonumber
\end{array} 
\label{betti}
\end{equation}
where the wedges between forms are understood.

The norm one $(3,0)$--form is given by 
\begin{equation}
\Psi = \alpha \wedge\beta\wedge \gamma = -  
dz\wedge dv\wedge du \,.
\end{equation}
It is obvious from (\ref{dal}) that this form is also closed and
therefore holomorphic.
Hence $\omega = \Psi$, and it follows from (\ref{W5Stro}) that
\begin{equation}
{\cal W}_5 = 0\,.
\end{equation}
It can also be explicitly checked that
\begin{equation}
J_0\wedge d J_0 = 0\,.
\end{equation}
Therefore also 
\begin{equation}
{\cal W}_4 = 0
\end{equation}
and hence (\ref{torsion}) is satisfied.

Having checked that the conditions on the torsion imposed by supersymmetry
are satisfied, we proceed to solve the supersymmetry conditions in the
gauge sector.  
We therefore compute the three--form field $H$, which we find to be
non--vanishing and given by
\begin{equation}
\begin{array}{rcl}
H &\equiv& \frac{i}{2} ({\bar \partial} - \partial) J_0 = -\frac{1}{4} 
\left(du -z dv\right) \wedge d {\bar z} \wedge d {\bar v} -
\frac14 \left(d {\bar u} - \bar z d\bar v\right) \wedge dz \wedge 
dv \\[3mm]
&=& \frac14\left(\bar \alpha \wedge \bar \beta \wedge \gamma + \alpha 
\wedge \beta \wedge \bar \gamma\right)\,.
\end{array}
\label{habc}
\end{equation}
We note that $H$ is entirely expressed in terms of $(2,1)$ and $(1,2)$--forms
which are not closed (cf. (\ref{betti})).
It is also interesting to note that the components of this 
tensor are proportional to the structure constants of the algebra.
Furthermore we find 
\begin{equation}
dJ_0 = \frac{i}{2} \left(\al \wedge \be \wedge \gabar 
- \ga \wedge \albar \wedge 
\bebar \right) \;,
\end{equation}
which differs from (\ref{habc}) by a relative sign.  
It then follows from (\ref{defw3}) that 
\begin{equation}
{\cal W}_3= \frac{i}{2} \, \al \wedge \be \wedge \gabar \;.
\end{equation}

The $H$--field also defines the torsion of the generalized connection
in the covariant derivative ${\cal D}$, see (\ref{covd}), from which
the generalized Riemann two--form ${\tilde R}^m\,_n$ can be
constructed.
The generalized connection is compatible with the metric and the 
complex structure is covariantly constant with respect to this 
connection:
\begin{equation}
{\cal D}_m g_{np} = 0\,, \quad {\cal D}_m {J_n}^p = 0\,.
\label{compat}
\end{equation}
The generalized Riemann 
two--form has the following non--zero components,
\begin{equation}
\begin{array}{rclcrcl}
{\tilde R}^z{}_z &=& d v \wedge d\bar v\,,  &\quad&{\tilde 
R}^z{}_v &=& d \bar v \wedge dz\,, \\[3mm]
{\tilde R}^v{}_z &=& -d v \wedge d\bar z\,,  &\quad&{\tilde 
R}^v{}_v &=& d z \wedge d\bar z\,, \\[3mm]
{\tilde R}^u{}_z &=& -z\,d v \wedge d\bar z\,,  &\quad&{\tilde 
R}^u{}_v &=& z\,d  v \wedge d\bar v + 2 z\, dz \wedge d\bar z\,, \\[3mm]
{\tilde R}^u{}_u &=& -d v \wedge d\bar v- dz \wedge d\bar z\,,
  &\quad& && 
\end{array}
\label{genRiemann}
\end{equation}
as well as the complex conjugate entries.
Note that there are no non--vanishing ${\tilde R}^a\,_{\bar b}$
two--form components which shows that the holonomy is contained in $U(3)$.
The generalized Ricci tensor is {\it not} vanishing and is given by
\begin{equation}
\tilde{R}_{z \bar z} = \tilde{R}_{\bar z z}= 
\tilde{R}_{v\bar v} = \tilde{R}_{\bar v v} 
= -1\,. 
\label{genRicci}
\end{equation}
Furthermore, it can be explicitly
checked that the generalized Ricci two--form is 
vanishing,
\begin{equation}
{\tilde R}_{mn}\,^{pq} J_{pq} = 0\,,
\label{genricci2form}
\end{equation}
which implies a further reduction of the holonomy group of the Iwasawa manifold
to be contained
in $SU(3)$ \cite{Gutowski:2002bc,strongKT}.
Therefore there exist two covariantly constant Killing spinors 
$\eta_{\pm}$ which are annihilated by the holonomy generators,
\begin{equation}
{\tilde R}_{mn}\,^{pq} \Gamma_{pq} \eta_{\pm} = 0\,.
\label{hol}
\end{equation}
In particular, by explicit computations it can be checked that the 
Iwasawa manifold endowed with the standard complex structure $J_0$ has $SU(2) 
\times U(1)$ holonomy.

In order to solve the
Bianchi identity (\ref{ddJ}) for $H$ 
we need to evaluate the four--forms $i\, \partial {\bar \partial} J_0$ 
and 
$tr (\tilde R \wedge \tilde R)$.
The first one is given by
\begin{equation}
\label{lhb}
 i \, \partial {\bar \partial} J_0 = {1\over 2}
 dz \wedge  dv\wedge  d{\bar z}\wedge 
d{\bar v} = \alpha \wedge \beta \wedge {\bar \alpha} \wedge {\bar \beta}\,,
\end{equation}
and is therefore proportional to the volume of the ${\mathbb T}^2
\times {\mathbb T}^2 $ base.
The other is identically zero
\begin{equation}
tr (\tilde R \wedge \tilde R) = 0\,.
\label{RR}
\end{equation}
This implies that a solution of the Bianchi identity will be now 
provided by a $(1,1)$--form $F$ such that $Tr(F \wedge F)$ is 
proportional to the volume of the base space and that $F_{mn} J^{mn} = 
0$.
In the following we will use an Abelian field strength configuration,
and not the more common $SU(n)$ configurations.
Let us therefore recall the origin of the terms in the Bianchi identity. 
The first and second Chern class of an $U(2)$ vector bundle $V$ is
defined from the following expansion of the full Chern class
$c(V)=1+c_1(V)+c_2(V)$, where
\begin{equation}
c(V)=\det \left({\bf 1 } + \frac{i}{2\pi} F^{ab}\right) \;,
\end{equation}
and where $a, b$ are the group indices of the $U(2)$ bundle, i.e. 
they refer to the matrix indices $1, 2$ of the element of $Lie U(2)$. 
Therefore one gets
\begin{eqnarray}
c_1(V) &=& \frac{i}{2\pi} tr F \;,\nonumber\\
c_2(V) &=& -\frac{1}{4\pi^2}
(F^{11}\wedge F^{22} - F^{21}\wedge F^{12})
= \frac{1}{8\pi^2}( tr F\wedge F - tr F \wedge tr F) \;.
\end{eqnarray}
Thus, for the actual anomaly term, we obtain
\begin{equation}
c_2(V)-\frac{1}{2}c_1^2(V)=\frac{1}{8\pi^2} tr F\wedge F \;.
\end{equation}
Note that in our case the left hand side of the Bianchi identity is
proportional to $\alpha \wedge \bar{\alpha} \wedge \beta \wedge
\bar{\beta}$ (cf. (\ref{lhb})). 
This entails two things. 
First, since on the one hand this form denotes 
the volume form of the base ${\mathbb T}^2
\times {\mathbb T}^2
$, whereas on the other hand it is 
nothing but $dH$ and hence it is also 
exact, there appears to be an immediate contradiction.
This apparent contradiction is,
however, avoided by noticing that no section exists
for a  nontrivial ${\mathbb T}^2$
fibration over the ${\mathbb T}^2 \times {\mathbb T}^2$
base, i.e. the base does
not lie embedded in the total space. 
Secondly, it is precisely this combination of four leg directions in
$dH$ which makes it possible to solve $dH\sim tr F\wedge F$ with just
an Abelian gauge field configuration.
This is so because in the Abelian case the field strength two--form $F$ is
closed, and $H^{1,1}(M)$ is precisely generated by $\alpha \wedge
\bar{\alpha} , \alpha \wedge \bar{\beta} , \beta \wedge \bar{\alpha} ,
\beta \wedge \bar{\beta} $ (cf. (\ref{betti})).

Specifically, let us pick a $U(1)$ subgroup of $U(2)$ in the
following way,
\begin{eqnarray}
F = \frac12 \, \left( \begin{array}{cc} {\cal F} & 0  \\ 0 & {\cal F} 
\end{array}\right) \;. 
\end{eqnarray}
The $(1,1)$--form $\cal F$ can be determined as follows.  Setting
\begin{equation}
{\cal F} = {\cal F}_{z \bar z} \, dz \wedge d{\bar z} 
+ {\cal F}_{v \bar v} \, dv \wedge d{\bar v}
+
{\cal F}_{z \bar v} \, dz \wedge d{\bar v} +
{\cal F}_{v \bar z} \, dv \wedge d{\bar z} 
\end{equation}
and requiring ${\cal F}$ to be real yields
\begin{equation}
{\cal F}_{z \bar z} = - {\cal F}^*_{z \bar z} \;,\quad
{\cal F}_{v \bar v} = - {\cal F}^*_{v \bar v} \;, \quad
{\cal F}_{z \bar v} = - {\cal F}^*_{v \bar z} \;.
\end{equation}
Demanding that ${\cal F}_{mn} J^{mn} = 
0$ yields $ {\cal F}_{z \bar z} = - {\cal F}_{v \bar v}$ in addition.
It follows that
\begin{equation}
tr F \wedge F = {\cal F} \wedge {\cal F} 
= - 2\, ( f^2 +| {\cal F}_{z \bar v}|^2 ) \, 
dz \wedge  dv\wedge  d{\bar z}\wedge 
d{\bar v}\,,
\end{equation}
where we set ${\cal F}_{z \bar z} \equiv i f$ with $f$ real.  
Setting $tr F \wedge F = 
- dH = 
-\frac12\, dz \wedge  dv\wedge  d{\bar z}\wedge 
d{\bar v}$ yields 
\begin{equation}
| {\cal F}_{z \bar v}|^2 = \frac{1}{4} -f^2 \;.
\end{equation}
Therefore 
\begin{equation}
{\cal F}_{z \bar v} = {\rm e}^{i \alpha}\, \sqrt{\frac{1}{4} -f^2} \;,
\end{equation}
and hence
\begin{equation}
{\cal F} = i f \, dz \wedge d{\bar z} - i f \, dv \wedge d{\bar v}
+ {\rm e}^{i \alpha}\, \sqrt{\frac{1}{4} -f^2}
 \, dz \wedge d{\bar v} -
 {\rm e}^{-i \alpha}\, \sqrt{\frac{1}{4} -f^2}
\, dv \wedge d{\bar z} \;.
\end{equation}
Next, we evaluate the Bianchi identity $d {\cal F}=0$.  
Equating the coefficients of the various $(2,1)$--forms yields
\begin{eqnarray}
i \partial_v f + \partial_z \Big({\rm e}^{-i \alpha}
\, \sqrt{\frac{1}{4} -f^2}
\Big) &=& 0 \;, \nonumber\\
i \partial_z f + \partial_v \Big({\rm e}^{i \alpha}
\, \sqrt{\frac{1}{4} -f^2}
\Big) &=& 0 \;.
\label{pf}
\end{eqnarray}
One solution to (\ref{pf}) is given by
\begin{equation}
{\cal F}_{z \bar v} = 0 \;\;\;,\;\; f = \pm \frac{1}{2} \;.
\end{equation}
Another solution is given by
\begin{equation}
{\cal F}_{z \bar v} = \frac{{\rm e}^{i \alpha}}{2}  \;\;\;,\;\;
f=0 \;\;\;,\;\;\; \alpha = {\rm constant} \;. 
\end{equation}

\subsection{The edge}

On the edge (\ref{line}) the complex structure is given by \cite{AGS}
\begin{equation}
J = - f^1 \wedge f^{2} + f^3 \wedge f^{4} - e^{5} \wedge e^{6} \;,
\end{equation}
and the associated $(1,0)$--forms are 
\begin{equation}
\kappa =   f^1 - i f^2 \,, \quad
\lambda =f^3 + i f^4 \,, \quad
\mu = e^5 - i e^6\,.
\end{equation}
These forms are related to the forms $\alpha, \beta$ and $\gamma$
introduced in (\ref{real}) by
\begin{equation}
\kappa = P {\bar \alpha}  \;\;,\;\;
 \lambda = P \beta \;\;,\;\; \mu = {\bar \gamma} \;,
\end{equation}
where $P \in SO(4)$.  
The norm one $(3,0)$--form $\Psi$ reads
\begin{equation}
\Psi = \kappa \wedge \lambda \wedge \mu \;.
\end{equation}
On the edge ${\cal W}_4=0$ \cite{AGS}.  
We now show that also ${\cal W}_5=0$, which implies that $\Psi$ is holomorphic.
Since $P$ is linear, it follows that
$d\Psi=P({\bar \alpha} \wedge
\beta)\wedge({\bar \alpha} \wedge {\bar \beta} )$.  Following \cite{AGS}, 
we decompose the two--forms into self-- and anti--selfdual
parts with respect to the four--dimensional Hodge star operator $\star_4$, i.e.
$\Lambda^2 D=
\Lambda^2_+ D \oplus \Lambda^2_- D$ (where $D = \oplus_{i=1}^4 e_i 
{\mathbb R}$).
Now ${\bar \alpha}  \wedge \beta$     is in $\Lambda^2_- D$
and ${\bar \alpha}  \wedge {\bar \beta} $ is in $\Lambda^2_+ D$.
Since $P = P_+ + P_-$ respects the decomposition, 
and wedging a form in $\Lambda^2_+ D$ with a form in $\Lambda^2_- D$
gives zero, $d\Psi=0$.

Computing $dJ$ and $H$ gives
\begin{eqnarray}
dJ &=& \frac{i}{2} \, \left( \alpha \wedge \beta \wedge {\bar \gamma} -
\gamma \wedge {\bar \alpha} \wedge {\bar \beta} \right) \;,
\nonumber\\[3mm]
H &=& \frac{1}{4} \, \left( \alpha \wedge \beta \wedge {\bar \gamma} +
\gamma \wedge {\bar \alpha} \wedge {\bar \beta} \right) \;.
\end{eqnarray}
Therefore we see that on the edge, the values for $dJ$ and for $H$ 
precisely equal the values obtained from $J_0$.  
This implies that the rest of the analysis given in the previous
subsection goes through also for the edge.
The holonomy with respect to the generalized connection is again
$SU(2) \times U(1)$, and the Bianchi identity (\ref{ddJ}) can again be
solved by an Abelian field strength.

Thus we have established that for any point in the moduli space of
complex structures the Iwasawa manifold is a consistent ${\cal N}=1$
supersymmetric heterotic background.

\section{Outlook}

Let us briefly describe a few directions for future research.  
First of all, it would be important to construct examples of complex,
conformally balanced six--dimensional spaces satisfying $2 {\cal W}_4
+ {\cal W}_5 =0$.
Finding such examples would enlarge the list of
non--K\"ahler manifolds representing consistent backgrounds for the
heterotic string provided that the conditions in the gauge sector
can also be met.
Second, it would be instructive to directly obtain the supersymmetry
conditions (\ref{ddagger})--(\ref{HJ}) from the ten--dimensional
heterotic action by rewriting the latter in terms of squares of
expressions which vanish for supersymmetric compactifications.
This would also provide a check of the equations of motion, which we
haven't performed in this paper.
It would be equally interesting to derive the ${\cal N}=1$
supersymmetry conditions (\ref{ddagger})--(\ref{HJ}) from the
minimization of a four--dimensional superpotential $W$, i.e. $W =
\partial_{\Phi} W =0$, where $\Phi$ denotes one of the background
fields.

The Iwasawa manifold which, as we showed, provides a consistent
background for the heterotic string, is an example of a half--flat
manifold and can therefore be lifted to a $G_2$ manifold by using the
flow equations (\ref{hitchin}) of Hitchin.
It would be interesting to make this explicit.

In \cite{Govindarajan:1987iz} it was shown that in the presence of
both an $H$--field and a suitably tuned gaugino condensate it is
possible to obtain supersymmetric backgrounds where the
four--dimensional spacetime part is an $AdS_4$ space.  
It would be very interesting to redo the analysis of
\cite{Strominger:1986uh} for the case that both an $H$--field and a
gaugino condensate are turned on and to reformulate the resulting
geometrical conditions on the internal manifolds in terms of the
mathematical classification of torsion described in section 2.

Another direction would be to discuss type II and M--theory compactifications
with fluxes in the mathematical framework of section 2
(cf. appendix A).

Finally let us comment on the
possibility of M--theory
compactifications on $G_2$ manifolds with boundaries and their
relation to six--dimensional heterotic string
backgrounds\footnote{M--theory compactifications on orbifolds to
four dimensions and their heterotic string interpretation were
discussed in 
\cite{Doran:2001ve,Doran:2002tg,Doran:2002iz}.} (cf. \cite{Curio:2000dw}).
As shown in \cite{Horava:1996qa} an M--theory background allows for a
ten--dimensional heterotic string interpretation if it possesses two
boundaries, namely the two ten--dimensional end of the
world planes on which the heterotic gauge degrees of freedom are
located.
Hence, a seven--dimensional $G_2$ manifold with two 
boundaries is expected to be associated with a particular ${\cal N}=1$
(non--perturbative) heterotic string background.
As a candidate for this scenario consider a $G_2$ manifold 
constructed as a warped
product $X_7= M \times I$, where $I\subset {\mathbb R}$
is a compact interval.
According to the Ho${\rm \check{r}}$ava--Witten picture, the 
radial $r$--direction would then correspond to the
eleventh M--theory dimension.  
The manifold $M$ would be the geometric heterotic background, and the
torsion could be due to a non--perturbative gaugino condensate.
Whether the associated Bianchi identity can be solved and whether
such a picture really survives remains to be seen.  
One may also study its possible relation to
heterotic coset space compactifications
\cite{Lust:1985be,Lust:1986ix,Castellani:1988rg}.

\bigskip \bigskip
\noindent
{\bf Acknowledgments}

\medskip

We would like to thank S. Chiossi, T. Friedrich and J. Louis
for very valuable discussions as well as 
S. Salamon and A. Strominger for very useful correspondence.
The work of G.L.C.\ is supported by the DFG.
Work supported in part by the European Community's Human Potential
Programme under contract HPRN--CT--2000--00131 Quantum Spacetime.

\vspace{1cm}

\noindent
{\bf Note added}

\noindent
As we were about to publish the paper, we received \cite{GLMW} which
has some overlap with the issues discussed in appendix A.

\appendix

\setcounter{equation}{0}

\section{$M$--theory compactifications}

In this appendix we will give
some examples of the occurrence of the five classes of torsion in 
string theory 
and in $M$--theory compactifications.

Some interesting subclasses of six--dimensional 
manifolds with $SU(3)$ structure arise in an indirect way as 
bases of seven--dimensional manifolds with $G_{2}$ holonomy 
(see also e.g. the discussion in \cite{Bilal:2001an}).
A $G_2$ structure on a a seven--dimensional manifold $X_7$ is
determined by a non--vanishing
3--form $\phi$ on $X_7$ given by
\begin{eqnarray}
\phi&=&e^1\wedge
e^3\wedge e^5+ e^1\wedge e^2\wedge e^7+e^3\wedge e^4\wedge
e^7\nonumber\\ &+& e^5\wedge e^6\wedge e^7-e^1\wedge e^4\wedge
e^6-e^2\wedge e^3\wedge e^6-e^2\wedge e^4\wedge e^5\, .
\end{eqnarray}
It is well known that
a necessary and sufficient condition for $G_2$ holonomy is that 
$\phi$ is closed as well as co--closed:
\begin{equation}
d\phi=0\, , \quad d\star\phi=0\, .
\end{equation}
This condition is equivalent to that $X_7$ admits a covariantly constant
Killing spinor $\eta$:
\begin{equation}
D_i\eta=0,\quad i=1,\dots ,7\, .
\end{equation}
Integrability of this equation leads to
\begin{equation}
R_{abij}\gamma^{ab}\eta=0\, ..\label{curvature}
\end{equation}
Since the torsion--free curvature satisfies
$R_{a[bcd]}=0$ one can conclude (by multiplying eq.(\ref{curvature})
by $\gamma^j$) that
\begin{equation}
R_{ij}\gamma^i\eta=0\, ,
\end{equation}
which implies that $R_{ij}=0$, i.e. $X_7$ is Ricci--flat.

In the following, we will discuss two types of $G_2$ structures,
namely dynamical and fibered $G_2$ structures \cite{chiossi}.

\subsection{Manifolds with dynamical $G_2$ structures}

As mentioned in section 2, a six--dimensional half--flat manifold $M$ can
be lifted as a warped product $X_7 = M \times I$ to a seven--dimensional
manifold with $G_2$ holonomy.
We now specialize to the case where $I$ is the half--interval
$I={\mathbb R}^+$, and the metric $g$ of $X_7$ 
is the conical metric $g=r^2\hat g+dr^2$.
The physical relevance of this metric comes from the fact that
chiral fermions in M--theory on a $G_2$ manifold
are located at those points where $X_7$ exhibits a
conical, codimension 7 singularity. In brane
language, these are precisely the points where D6--branes intersect each other
in a supersymmetric way.

The uplift of the six--dimensional (hatted) $SU(3)$--structure to  
seven dimensions follows from the form of the conical metric and is given by
\begin{equation}
J=r^2 \hat J,\quad \psi_+=r^3 \hat\psi_+,\quad \psi_-=r^3
\hat\psi_-\,.
\end{equation}
Following section 2, the $G_2$ structure is defined by
$\phi = J \wedge dr + \psi_{+}$. The Hitchin flow equations (\ref{hitchin})
reduce to
\begin{equation}
\hat d \hat \psi_-=-2~\hat J\wedge\hat  J\, ,\qquad
\hat d \hat J=3~\hat \psi_+\, .\label{weaksu3}
\end{equation}
This implies a further restriction of the allowed torsion to
\begin{equation}
\tau\in{\cal W}_1^-\ .
\end{equation}
Therefore $M$ is nearly--K\"ahler (cf. (\ref{nearlyKaehler})).  $M$ 
is also referred to as having
weak $SU(3)$ holonomy.

Examples of nearly--K\"ahler manifolds are the homogeneous spaces
$M=Sp(4)/(SU(2)\times U(1))$, 
$M=SU(3)/(U(1)\times U(1))$ and
$M=SU(2)^2/SU(2)$.
The metrics of the three associated
$G_2$ manifolds are well known. For $M=SU(2)^3/SU(2)$
the seven--space $X_7$ is a ${\mathbb R}^4$ bundle over $S^3$,
whereas for $M=Sp(4)/(SU(2)\times U(1))$ or $M=SU(3)/U(1)^2$
the corresponding non--compact $G_2$ manifolds are a ${\mathbb R}^3$
bundle over a quaternionic space $Q$, with $Q=S^4$ or $Q={\mathbb 
C}{\mathbb P}^2$
respectively.
Note that there also exist more general quaternionic spaces with less
isometries which can be used to construct $G_2$ manifolds.
In these cases $M$ will not be anymore a homogeneous coset
space, but nevertheless one can still construct its 
weak $SU(3)$ structure \cite{Behrndt:2002xm}.

For the homogeneous coset spaces the relevant torsion tensor can be
explicitly constructed. Namely $\tau$ and also $\psi_-$ are directly
proportional to the structure constants $f_{abc}$ of the coset space.
These structure constants appear in the connection which possesses
the $SU(3)$ structure discussed above. This connection is non--Ricci
flat.   However,
there also exists another connection, with torsion again given
by the structure constants $f_{abc}$, which is Ricci flat 
\cite{Lust:1986ix}.
A priori it is not clear whether the manifolds with weak $SU(3)$ 
holonomy like the coset spaces listed above
can be used as geometric background spaces in string theory.
These are certainly not to be used as type IIA backgrounds, since
they do not follow from a circle reduction.

\subsection{Type IIA compactifications: manifolds with fibered $G_2$
structures}

The well--known case of Calabi--Yau compactifications of type IIA
leads to an effective four--dimensional ${\cal N}=2$
supergravity theory.
However one can turn on additional background fields in order to achieve a 
partial supersymmetry breaking from ${\cal N}=2$ to ${\cal N}=1$
supersymmetry. 
One possibility is given by turning on appropriate
Ramond 2--form H--fluxes through 2--cycles of the Calabi--Yau space.
Another way is given by wrapping D6--branes around supersymmetric
3--cycles of the Calabi--Yau space.
Examples are given by the resolved conifold with
$N$ units of 2--form H--flux through the $S^2$, resp. 
the deformed conifold with $N$ wrapped D6--branes around the $S^3$
\cite{Vafa:2000wi}.
The resulting effective field theory possesses precisely ${\cal N}=1$
supersymmetry. Both cases have in common that the internal background fields
are not given only by the 6--dimensional metric but also by some
additional `matter' fields, the Ramond 1--form potential and the
dilaton field. In fact, turning on these additional background fields
has a strong backreaction on the internal geometry, such that the modified
metric of the six--dimensional space $M$ does not belong
any more to a Calabi--Yau manifold, but instead 
to a non--Ricci flat space with a
certain $SU(3)$ structure.

Let us consider the case 
where the additional type IIA matter background fields
become purely geometric when uplifting type IIA to M--theory.
Asking for ${\cal N}=1$ supersymmetry in four space--time dimensions,
the 7--dimensional M--theory background space $X_7$ must have $G_2$
holonomy.
E.g., for the 6--dimensional conifold with additional 2--form flux or
with wrapped D6--branes this leads to a particular non--compact $G_2$
manifold, which topologically is an $R^4$ bundle over $S^3$
\cite{Acharya:2000gb,Atiyah:2000zz,Brandhuber:2001yi,Cvetic:2001kp}.
For the IIA reduction one 
identifies a particular compact circle $S^1_{11}$
inside $X_7$. This circle is usually non--trivially fibered
over a six--dimensional base $M$ which then serves as the non--Ricci flat
geometric background of the corresponding IIA superstring theory.
In M--theory language the D6--branes correspond to co--dimension 4
A-D-E singularities of $X_7$. Moreover the brane intersection points,
where the chiral fermions are localized, are given in terms of co--dimension
7 singularities of $X_7$.

In order to determine the $SU(3)$ structure of the type IIA
background $M$, we consider the circle fibration \cite{chiossi}
\begin{equation}
\pi:\quad X_7\rightarrow M\, ,
\end{equation}
endowed with a metric of the form
\begin{equation}
g=\alpha\otimes \alpha+\pi^*\hat g\, ,
\end{equation}
where $\hat g$ is the metric of $M$, and $d\alpha=\pi^*\rho$
with $\rho$ some 2--form on $M$. 
$\rho$ corresponds
to the Ramond 2--form field strength background and hence $\alpha$
is the Ramond 1--form gauge potential.
A $G_2$ structure of $X_7$ can be introduced as
\begin{equation}
\phi = J \wedge \alpha + \psi_+\,
\label{phiIIA}
\end{equation}
and it satisfies the following two relations (and
omitting the pullback operator):
\begin{eqnarray}
d\phi&=& dJ \,\wedge\alpha+d\psi_+ +J\,\wedge\rho\, ,\nonumber\\
d\star\phi&=& d\psi_- \wedge\alpha+J\wedge dJ-\psi_-\wedge \rho\, .
\end{eqnarray}
Now,  requiring that $X_7$ is Ricci flat and $g$ has $G_2$ holonomy 
implies that $d\phi=d\star\phi=0$.
This may be achieved by demanding that the almost complex structure is 
closed,
\begin{equation}
d J = 0\,,
\end{equation}
and that the $(3,0)$--form $\Psi$ satisfies
\begin{equation}
d\psi_-=0\, ,\qquad d\psi_+=-J\wedge\rho\,.
\end{equation}
Moreover the remaining condition $\rho \wedge \psi_- =0$ implies that
\begin{equation}
\rho \in \Lambda^{(1,1)}_0\,,
\label{rho}
\end{equation}
which means that $\rho$ is a $(1,1)$--form which cannot 
contain any piece proportional to $J$.
Therefore the torsion is in 
\begin{equation}
\tau\in{\cal W}_2^-\, .
\end{equation}
So the type IIA space $M$ is an {\it almost--K\"ahler manifold} 
(cf. (\ref{almostKaehler})).

Note that for case of Ramond 2--form flux through the 2--cycle of the
resolved conifold, the non--closure of the 3--form $\psi_+$ of $M$ corresponds
to a non--vanishing NS 4--form flux as discussed in
\cite{Vafa:2000wi} (see also the remarks in section 5 of 
\cite{Curio:2001dz}).
So turning on the non--geometrical Ramond 2--form flux induces as a backreaction
on the six--dimensional internal space a geometrical 4--form flux, which
is indeed needed to preserve ${\cal N}=1$ supersymmetry. In the effective
${\cal N}=1$ superpotential the non--geometrical 2--form flux as well as the
geometrical 4--form
flux are present, as they balance each other in the supersymmetric
ground state of the scalar potential.
In the type IIB mirror description both fluxes are non--geometrical,
they correspond to a Ramond plus NS 3--form flux on the mirror geometry,
which is still a Calabi--Yau space in type IIB.
In the type IIB superpotential the contribution of the Ramond 3--form
flux is balanced by the NS 3--form flux in order to get a supersymmetric
groundstate.
Therefore this
means that type IIA on the non--Calabi Yau space $M$ plus 2--form
and 4--form background fluxes
is mirror to IIB on a Calabi--Yau with appropriate 3--form H--fluxes.

\section{Supersymmetric compactifications of the heterotic theory}

In this appendix we review the derivation of the conditions 
leading to ${\cal N}=1$ spacetime supersymmetry when
compactifying heterotic string theory on a six--dimensional manifold
\cite{Candelas:1985en,Strominger:1986uh}.

The supersymmetry transformations of the fermionic fields in the 
heterotic theory are given by \cite{Bergshoeff:1982um,Chapline:1983ww,
Horava:1996vs}:
\begin{subequations}
\begin{eqnarray}
\delta \psi_{M} &=& \nabla_{M}\epsilon + \frac{\sqrt{2}}{32} \, 
\varphi^{-3/4} \left( \Gamma_{M} \Gamma^{NPQ} - 12 \delta_{M}^N 
\Gamma^{PQ} \right) \epsilon \, H_{NPQ}+ \nonumber\\
&-&\frac{1}{256}\, \left(
\Gamma_{M} \Gamma^{NPQ} - 8 \delta_{M}^N 
\Gamma^{PQ} \right) \epsilon \, \bar\chi \Gamma_{NPQ} \chi + \ldots 
\,, \label{gravitino}\\
\delta \chi &=& - \frac14 \, \varphi^{-3/8}\,\Gamma^{MN}\epsilon 
\,F_{MN} + \ldots\,,
\label{gaugino}\\
\delta \lambda &=& -\frac{3\sqrt{2}}{8} \, \varphi^{-1} \nabla\!\!\!\!\slash 
\varphi \, \epsilon + \frac18 \varphi^{-3/4}\,\Gamma^{MNP} \epsilon \, 
H_{MNP} + 
\frac{\sqrt{2}}{384}\, \left(\bar\chi \Gamma_{NPQ} \chi \right)
\, \Gamma^{NPQ} \, \epsilon + \ldots 
\label{dilatino}
\end{eqnarray}
\label{susy}
\end{subequations}

\noindent
where the dots stand for other Fermi terms.
The capital indices $M,N$, label the ten--dimensional coordinates and 
$\psi_{M}$ is the gravitino, $\chi$ the gluino, $\lambda$ the dilatino, 
$\phi$ the dilaton, $F$ a Yang--Mills field--strength and $H$ the three--form field that 
satisfies the Bianchi identity
\begin{equation}
d H = \hbox{tr} \tilde R\wedge \tilde R - 
\hbox{tr} F\wedge F \,,
\label{Bianchi}
\end{equation}
where $\tilde R$ denotes the Riemann tensor computed with respect to a 
generalized connection that includes torsion terms.
We have reported here only the relevant terms for both 
compactifications with $H$ fluxes and/or gaugino condensates 
$\langle \bar \chi 
\Gamma^{MNP} \chi\rangle $.

We are interested here in compactifications where the metric reduces 
to:
\begin{equation}
ds^2 = g_{MN}^0 dx^M \otimes dx^N =
{\rm e}^{2\Delta(y)} \left( dx^\mu \otimes dx^\nu \, \hat g_{\mu\nu}(x)
+ dy^m \otimes dy^n \, \hat g_{mn}(y)\right)\,,
\label{metric2}
\end{equation}
where $\Delta$ is the so--called warp factor, depending only on the 
internal coordinates, while the four--dimensional metric $\hat g_{\mu\nu}$ is 
that of a maximally symmetric spacetime.
We will also require the dilaton to depend only on the 
internal coordinates $\phi = \phi(y)$, the $F$ and $H$ fields and the 
gaugino condensate to acquire expectation values only on the internal 
manifold.
We also note that the Killing spinor $\epsilon$ will split into a 
four--dimensional part $\varepsilon(x)$ and a six--dimensional one 
$\eta(y)$.

In the following we will not consider gaugino condensates.
A much simpler formulation of the 
supersymmetry equations (\ref{susy})
can then be obtained by rescaling them by appropriate 
powers of the dilaton field, such as to make them dilaton free.
The transformations of the relevant fields are given 
by 
\begin{equation}
\begin{array}{rclcrcl}
\varphi &=& {\rm e}^{-8/3 \phi} \,,&\quad&
g_{MN} & = &  {\rm e}^{-2 \phi} g_{MN}^0\,, \\[2mm]
\lambda & = & \frac{1}{\sqrt2} \, {\rm e}^{\phi/2} \lambda^0\,, &\quad&
\psi_{M} & = & {\rm e}^{-\phi/2} \left(\psi_{M}^0 - \frac{\sqrt{2}}{4}
\Gamma_{M}^0 \lambda^0\right)\,, \\[3mm]
\epsilon &  = & {\rm e}^{-\phi/2}\epsilon^0\,, &\quad& H_{MNP} &=& 
\frac{3}{\sqrt2}\,H_{MNP}^0\,,\\[3mm]
\chi &  = & {\rm e}^{\phi/2}\chi^0\,, &\quad& F_{MN} &=& F_{MN}^0\,,
\end{array}
\label{Hrescaling}
\end{equation}
where the quantities with a superscript $0$ refer to the ones in (\ref{susy}). 
After these rescalings, the equations (\ref{susy}) can be rewritten as
\begin{subequations}
\begin{eqnarray}
\delta \psi_{M} &=& \nabla_{M}\epsilon - \frac14 H_{M} \epsilon \,, 
\label{gravitinor2}\\
\delta \chi &=& - \frac14 \,\Gamma^{MN}\epsilon 
\,F_{MN}\,,
\label{gauginor2}\\
\delta \lambda &=&  \nabla\!\!\!\!\slash \phi + \frac{1}{24} \, 
H\, \epsilon  \,,
\label{dilatinor2}
\end{eqnarray}
\label{susyr2}
\end{subequations}

\noindent
where we further defined $H \equiv \Gamma^{MNP} H_{MNP}$, $H_{M} 
\equiv H_{MNP} \,\Gamma^{NP}$ and the covariant derivative $\nabla$ is 
constructed from the rescaled metric.

In the setup described above, 
once the variations of the Fermi fields are set to (\ref{susyr2}), 
the only way 
left to preserve some supersymmetry using a maximally symmetric 
four--dimensional space is that such space has vanishing cosmological 
constant \cite{Strominger:1986uh}
(in the case of both a gaugino condensate and an $H$--form flux, on the other
hand, it is possible to have a non--vanishing cosmological 
constant \cite{Govindarajan:1987iz}).
A short proof for this is given by the analysis of the integrability of the 
gravitino transformation  rule on the four--dimensional part
\begin{equation}
\delta \psi_{\mu}= \nabla_{\mu} \varepsilon = 0\,.
\label{psimu}
\end{equation}
As said above, the covariant derivative is built in terms of the 
rescaled metric.
The same formula in terms of the $\hat g_{\mu\nu}$ connection reads (here 
the hat denotes quantities computed from $\hat g_{\mu\nu}$)
\begin{equation}
\hat\nabla_{\mu} \varepsilon+ \frac12 {\Gamma_{\mu}}\Gamma^m \partial_{m} 
\,\log \left(\Delta -\phi\right)\,\varepsilon = 0\,,
\label{hatn}
\end{equation}
whose consequences, for a compact six--dimensional manifold, are a 
condition relating the dilaton profile to the warp factor
\begin{equation}
\Delta(y) = \phi(y) + {\rm constant}\,,
\label{dilatwarp}
\end{equation}
as well as the flatness of the spacetime 
\begin{equation}
\hat R = 0\,.
\label{Ricci4d}
\end{equation}
The space has been therefore simplified to the warped product of a Minkowski 
four--dimensional spacetime and an internal six--dimensional 
one, which we are now 
going to determine.

The same equation (\ref{gravitinor2}) on the internal directions tells us 
that there must be a six--dimensional 
Killing spinor which is covariantly constant with respect to the 
covariant derivative
\begin{equation}
{\cal D}_{m} \equiv \partial_{m} + \frac14 \left({\omega_{m}}^{np} - 
{H_m}^{np}\right) \Gamma_{np}\,.
\label{covd2}
\end{equation}
This shows that we have introduced a torsion term in the connection 
proportional to the three--form flux.
If the holonomy of the space reduces to $SU(3)$,
there are two Weyl spinors which are covariantly constant with respect 
to this generalized connection
\begin{equation}
{\cal D} \, \eta_{\pm} = 0 \,,
\label{killspin}
\end{equation}
one with positive $\eta_+$ and one with negative chirality $\eta_-$.
They can be chosen to be normalized to one
\begin{equation}
\eta_{\pm}^\dagger \eta_{\pm} = 1\,.
\label{normal}
\end{equation}
Two important objects can be constructed from these spinors 
\cite{Strominger:1986uh}: the almost complex structure $J$ and a 
$(3,0)$--form $\omega$ which will be shown to be holomorphic.
It can be proved that the manifold is complex by defining 
\begin{equation}
{J_{m}}^n \equiv i \, \eta^\dagger_{+} {\Gamma_m}^n\eta_+\,,
\label{structure}
\end{equation}
which, of course, satisfies
\begin{equation}
{J_m}^n {J_n}^p = - {\delta_m}^p\,,
\label{Jsquare}
\end{equation}
as can be seen by using the standard Fierz identity, plus some gamma 
algebra.
This almost complex structure is covariantly constant with respect to the 
generalized connection
\begin{equation}
{\cal D}_m {J_n}^p = \nabla_m {J_n}^p - H_{sm}{}^p {J_n}^s - {H^s}_{mn} 
{J_s}^p = 0\,.
\label{DJ0}
\end{equation}
The fact that $J$ is really a complex structure follows from the 
analysis of the Nijenhuis tensor ${N_{mn}}^p  \equiv {J_m}^q 
{J_{[n}}^p{}_{,q]} - {J_n}^q {J_{[m}}^p{}_{,q]}$.
Using the covariant constancy of $J$, it can be recast in the form
\begin{equation}
N_{mnp} = H_{mnp} - 3\, {J_{[m}}^q {J_n}^r H_{p]qr}\,,
\label{Nij}
\end{equation}
which can be proven to be zero using the equation for the dilatino 
(\ref{dilatinor2}) and again some gamma algebra.
This allows then to introduce complex coordinates and hence 
holomorphic $a,b,c$ and antiholomorphic $\bar a, \bar b, \bar c$ indices.
It also follows from (\ref{structure}) by some more gamma algebra that 
the metric is hermitian
\begin{equation}
g_{mn} = {J_m}^p {J_n}^q g_{pq}\,.
\label{hermiticity}
\end{equation}
Moreover, we can associate to $J$ the following two--form
\begin{equation}
J = \frac12 \,{J_n}^p \,g_{pm}\, dx^n \wedge dx^m = i\, g_{a\bar b} 
dz^a \wedge d\bar z^{\bar b}\,,
\label{Jform}
\end{equation}
which satisfies the duality relation
\begin{equation}
\star J = \frac12 \, J \wedge J\,.
\label{starJ}
\end{equation}
This can be verified using once more the Fierz identity, some 
gamma algebra and the normalization of the Killing spinor.
The definition for the Hodge dual used is 
\begin{equation}
\star \Phi_{(p)}= \frac{1}{p!(6-p)!}\, dx^{m_1} \wedge \ldots 
dx^{m_{6-p}} \epsilon_{m_1\ldots m_{6-p}}{}^{m_{6-p+1}\ldots m_6} 
\Phi_{m_{6-p+1}\ldots m_6}\,.
\label{dual}
\end{equation}

Using the vanishing of the Nijenhuis tensor in (\ref{Nij}), the 
covariant constancy of $J$ (\ref{DJ0}) and the properties of the 
complex structure, it can be shown that the three--from $H$ is
expressed in terms of the complex structure by
\begin{equation}
H = \frac{i}{2} \,\left(\bar\partial - \partial\right) J\,,
\label{HJ2}
\end{equation}
where the standard decomposition of the exterior derivative in 
holomorphic $\partial$ and antiholomorphic $\bar \partial$ components 
has been used.
The Bianchi identity for $H$ follows then in terms of $J$ as
\begin{equation}
dH = i \, \partial \bar \partial J = 
\hbox{tr} \, \tilde R \wedge 
\tilde R - \hbox{tr} F \wedge F\,.
\label{ddJ2}
\end{equation}

Not all the information of the supersymmetry equations (\ref{susyr2}) is 
contained in the previous relations.
The dilatino equation can be recast into a relation between the dilaton 
and the complex structure
\begin{equation}
8\, i \left( \bar \partial - \partial\right)\, \phi = dx^n \nabla_p {J_n}^p = \star d \star
J\,.
\label{dphi}
\end{equation}
A holomorphic (3,0)--form can be defined 
as follows,
\begin{equation}
\omega \equiv {\rm e}^{8 \phi}\, \eta_-^T \Gamma_{abc} \eta_- \, 
dz^a \wedge dz^b \wedge  dz^c\,.
\label{omega}
\end{equation}
The fact that this form is holomorphic follows from 
(\ref{gravitinor2}), (\ref{dphi}) and some more algebra.
The key feature is that its norm is determined by the 
dilaton, up to a constant factor,
\begin{equation}
||\omega|| = \sqrt{\omega_{abc} \;\omega_{\bar a \bar b \bar c} \;g^{a\bar 
a}\, g^{b \bar b}\, g^{c \bar c}}={\rm e}^{8 \,(\phi-\phi_0)}\,.
\label{norm}
\end{equation}
As a consequence, the relation (\ref{dphi}) 
between the dilaton and the complex 
structure becomes a purely geometrical relation between the complex 
structure and the holomorphic (3,0)--form\footnote{This equation 
differs by a sign from \cite{Strominger:1986uh}, as also later 
corrected in \cite{Strominger:1990et}.}:
\begin{equation}
i \left( \bar \partial - \partial\right)\, \log ||\omega|| = \star d \star J\,.
\label{ddagger2}
\end{equation}

The remaining gaugino equation (\ref{gauginor2}) becomes just a 
constraint for the two--form field--strength.
In particular, from $\eta^\dagger_+ \Gamma_{mn} \delta \chi = 0$, 
one obtains that $F$ must be a $(1,1)$--form further subjected to
\begin{equation}
F_{mn} J^{mn}  =0\,,
\label{FJ2}
\end{equation}
which is a consequence of the contraction of (\ref{gauginor2}) with $\eta^\dagger_+$.


\providecommand{\href}[2]{#2}\begingroup\raggedright\endgroup

\end{document}